\documentclass[showpacs,preprint,preprintnumbers,amsmath,amssymb]{revtex4}
\usepackage{graphicx}
\usepackage[dvips]{color}

\begin{document}

\title{Loading Bose condensed atoms into the ground state of an
optical lattice}

\author{P S. Julienne$^{\,1}$, C. J.Williams$^{\,1}$, Y. B.
Band$^{\,2}$, Marek Trippenbach$^{\,3}$}

\affiliation{${}^{1}$ Atomic Physics Division, A267 Physics,
National
Institute of Standards and Technology, Gaithersburg, MD 20899 \\
${}^{2}$ Department of Chemistry, Ben-Gurion University of the
Negev,
Beer-Sheva, Israel 84105 \\
$^{\,3}$ Institute of Experimental Physics, Optics Division,
Warsaw
University, ul.~Ho\.{z}a 69, Warsaw 00-681, Poland \\
}

\begin{abstract}
We optimize the turning on of a one-dimensional optical potential,
$V_L(x,t) = S(t) V_0 \cos^2(kx)$ to obtain the optimal turn-on
function $S(t)$ so as to load a Bose-Einstein condensate into the
ground state of the optical lattice of depth $V_0$.  Specifically, we
minimize interband excitations at the end of the turn-on of the
optical potential at the final ramp time $t_r$, where $S(t_r) = 1$,
given that $S(0) = 0$.  Detailed numerical calculations confirm that a
simple unit cell model is an excellent approximation when the turn-on
time $t_r$ is long compared with the inverse of the band excitation
frequency and short in comparison with nonlinear time $\hbar/\mu$
where $\mu$ is the chemical potential of the condensate.  We
demonstrate using the Gross-Pitaevskii equation with an optimal 
turn-on function $S(t)$ that the ground state
of the optical lattice can be loaded with very little excitation even
for times $t_r$ on the order of the inverse band excitation frequency.
\end{abstract}


\maketitle

\section{Introduction}

A number of competing schemes for implementing quantum information
and quantum computing are being explored; five physical systems
have been proposed as quantum logic gates: ion traps
\cite{Cirac&Zoller}, high-Q optical cavities
\cite{Brune,Turchette}, nuclear magnetic resonance systems
\cite{Gershenfeld}, solid-state qubits (semiconductor quantum-dot
and Josephson-junction devices) \cite{DiVi,supercond} and
ultra-cold neutral atoms in optical lattices
\cite{Brennen_99,Jaksch_99}.  The proposals for using atoms in
optical lattices can be implemented by first loading Bose
condensed atoms that are held by a weak magnetic field into
an optical lattice by gradually turning on the optical potential
to its desired strength. Upon increasing the intensity of the
optical lattice to a critical intensity, the Bose-Einstein
condensate (BEC) will undergo a quantum phase transition from a
superfluid state to a Mott-insulator state \cite{Jaksch_98}.  One
thereby can obtain one atom per lattice site in the ground state of
the system; these atoms can serve as qubits.  This suggestion of
preparing a Mott-insulator state has recently led to a seminal
experiment \cite{Greiner} wherein the quantum phase transition was
observed.  In principle, starting with a BEC in a trap and turning
on an optical lattice of sufficient well depth in a sufficiently
adiabatic manner, prepares the Mott-insulator state.  In practice,
it is easy to turn on the optical lattice adiabatically with
respect to band excitation (excitation from one band to another);
however, it is substantially more difficult to turn on the optical
lattice adiabatically with respect to quasi-momentum excitation.
The second, more stringent form of adiabaticity requires that the
optical lattice be switched on slowly with respect to mean-field
interactions and tunneling dynamics between optical lattice sites,
and hence typically requires milliseconds \cite{Band_02}.  We
refer to the first form of adiabaticity as `{\em inter}band
adiabaticity' and the second form as `{\em intra}band
adiabaticity'.  The intraband adiabaticity condition has been
demonstrated in one-dimensional lattices by Orzel {\it et
al.}~\cite{Orzel} and ultimately led to the pioneering
experimental demonstration of the Mott-insulator transition
\cite{Greiner}.

The goal of this paper is to investigate the loading of atoms into an
optical lattice in as short a time as possible, while maintaining
interband adiabaticity, so as to obtain maximal atomic population
in the ground state band with minimal band excitation.  We consider 
experiments of the type described in Ref.~\cite{Denschlag02}, which 
measured the interband nonadiabaticity.  We show how to
optimize the turning on of a one-dimensional (1D) optical lattice so
as to minimize the interband nonadiabaticity.  We carry out
calculations  using the Gross-Pitaevskii (GP)
equation, as well as simplified models.  We show that an optimized
lattice turn-on results in very
low nonadiabaticity even for loading times $t_r$ comparable to
$1/\Delta \omega$, where $\hbar \Delta \omega$ is the excitation
energy of the first band that can be excited.  On such timescales, 
the nonlinear term in the GP equation 
is small and has little effect on the dynamics.

Section II discusses the theoretical models used, and
describes how to determine the optimized turn-on of the lattice.  It
also describes a very simple unit cell model that is in excellent
agreement with the full calculations.  Section III describes our
results and compares to experimental data.  A final Section summarizes
our conclusions.

\section{Theory}

Our theoretical investigation was stimulated by experimental work to
load a BEC into a one-dimensional optical lattice so as to create the
ground state of atoms in the optical lattice \cite{Denschlag02}.  We
first briefly describe the optical potential and its turn-on, then
consider the mean-field description of the process, develop the simple
unit-cell model, and finally describe the optimization procedure used.

\subsection{The Optical Potential}

\begin{figure}
   \centering
   \includegraphics[scale=0.33,angle=0]{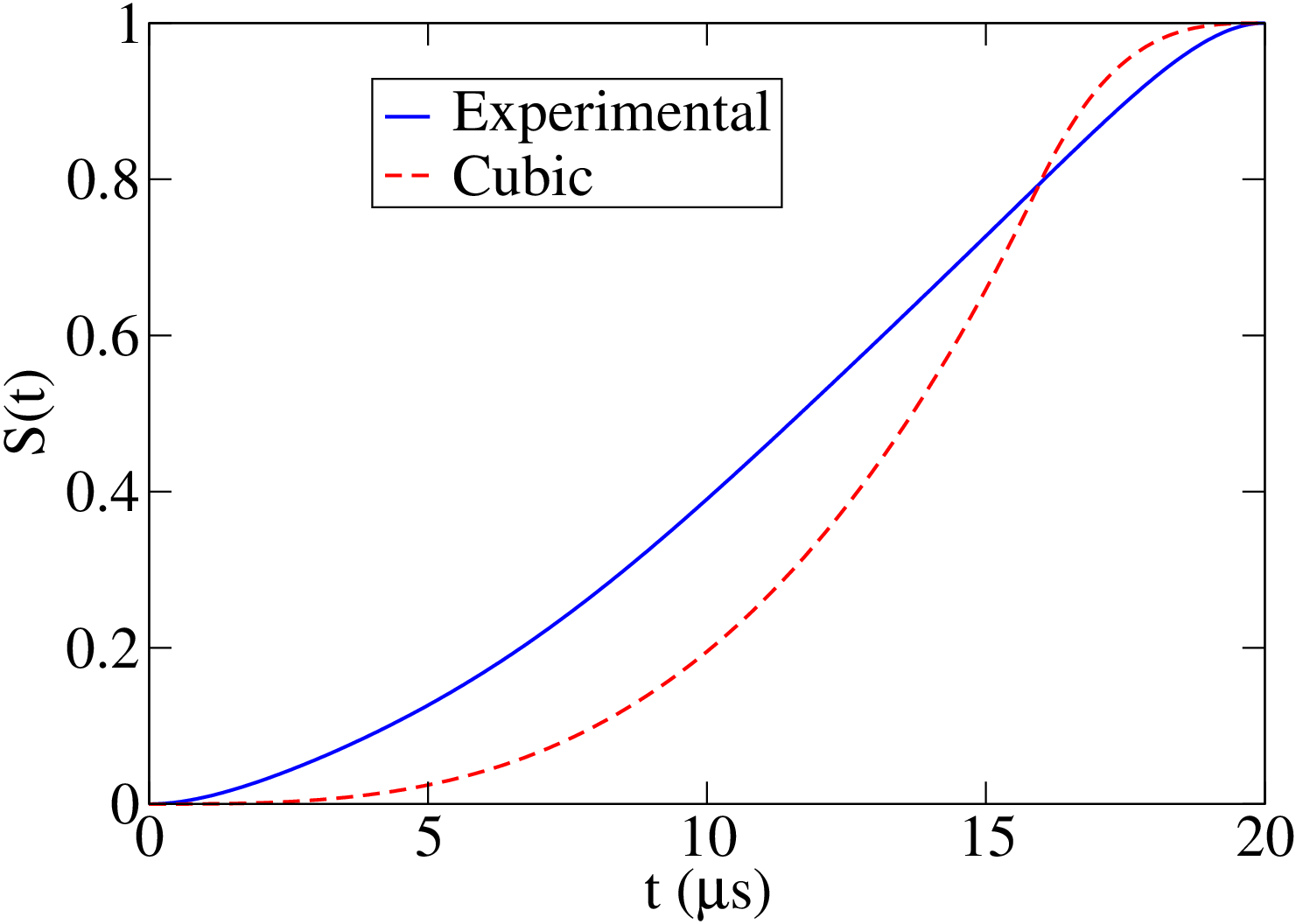}
   \caption{Fit to the experimental ramp function $S(t)$ used in
   Ref.~\cite{Denschlag02}, and piece-wise cubic $S(t)$ with $t_r =
   20\,\mu$s.  In both functions $t_r = 20\,\mu$s. \\}
   \label{fig.ramp}
\end{figure}

We start with a condensate of $N$ atoms in a trap with frequencies
$\nu_x$, $\nu_y$, and $\nu_z$.  At time $t=0$,
a 1D optical lattice with potential
\begin{equation}
V_L(x,t) = -S(t) V_0 \cos^2(kx) \label{Eq.1}
\end{equation}
is turned on along the $x$ direction.  The lattice periodicity equals
$\pi/k$,
where $k = \omega_L/c$ and $\omega_L/2\pi$ are the wave vector and
frequency of the laser beams that form the optical lattice.  The
lattice beams are sufficiently far-detuned so that spontaneous
emission is negligible on the time scale of the dynamics described
here.  The parameter $V_0$ is the final lattice depth after completion
of the turn on process at ramp time $t_r$.  The ramp turn-on function
$S(t)$ has only two constrains: $S(0)=0$ and $S(t_r)=1$.  For
diagnostic
purposes the lattice is held on with fixed depth $V_0$ until time $t_f
= t_r + t_{{\mathrm{hold}}}$, then turned off to allow for condensate
expansion and imaging.  Fig.~\ref{fig.ramp} shows
two examples of ramp functions with $t_r = 20\,\mu$s: a piece-wise
cubic
function with function and derivative matched at $t_m = 16\, \mu$s
and a fit to the experimental ramp function used in
Ref.~\cite{Denschlag02}.

\subsection{Mean-field Description}
\label{sec.mf}

The GP equation for the BEC wave function $\Psi({\mathbf r},t)$
(the order parameter of the condensate) is
\begin{equation}
i\hbar \frac{\partial \Psi}{\partial
t}=-\frac{\hbar^2}{2m}\nabla^2 \Psi + V({\mathbf r},t)\Psi
+\frac{4\pi\hbar^2 a}{m}|\Psi|^2\Psi \,, \label{Eq.2}
\end{equation}
where $V({\mathbf r},t)=V_{\mathrm{trap}}({\mathbf r})$ for times
$t < 0$ and $V({\mathbf r},t)=V_L(x,t)$ for $t \ge 0$, and the
nonlinear term is proportional to the two-body $s$-wave scattering
length $a$, and $m$ is the atomic mass.  We assume the trap is
turned off at $t=0$.  The lattice turn on time $t_r$ is assumed to be very 
short compared to the time needed for the condensate to expand
significantly.

The initial condensate wave function at $t=0$ is that for a trap
with time-invariant potential $V_{\mathrm{trap}}({\mathbf r})$.
Although the momentum distribution of the initial condensate wave
function $\Psi({\mathbf r},0)$ is sharply peaked close to zero,
the final condensate wave function $\Psi({\mathbf r},t_r)$ in the
lattice has developed momentum components peaked near 0, $\pm 2
\hbar k$, $\pm 4 \hbar k$, \ldots \,.  These components appear  even if 
only the ground state band of the lattice is occupied.  Consequently, the
subsequent free evolution of the condensate after $t_f$ results in
the further physical splitting of the condensate wave packet into
spatially separate parts having central momenta $2n \hbar k$,
where $n = 0, \pm 1$, \ldots \,.  The experiment consists of
measuring the fraction $p_{nk}(t_f)$ of the initial total number of
condensate atoms $N$ that appear in the $n$th wave packet.  As we
describe below, the signature of nonadiabatic excitation during
the turn-on of the ramp is oscillatory behavior of $p_{nk}(t_f)$
with varying hold times $t_{{\mathrm{hold}}}$.

The central goal of this paper is to determine the optimized ramp
function $S_{\mathrm{opt}}(t)$ for a given $t_r$ and $V_0$, i.e.,
the function $S(t)$ which leads to minimal oscillatory amplitude
of $p_{nk}(t_f)$, and hence minimal excitation of the final
condensate wave packet at time $t_r$ (see Eq.~(\ref{Eq.7})).  Our
optimization will be carried out using the GP equation.  We do not
require that our system remain adiabatic during the whole time
interval between $t=0$ and $t_r$, but only that the inter-band
excitation be as small as possible at the final time $t_r$.  It should 
be noted that the mean-field treatment used here can not be used 
upon increasing the
optical lattice potential to the point where the BEC undergoes a
quantum phase transition from its BEC-like superfluid state to a
Mott-insulator state.  Once the number of atoms in each optical
potential well becomes small (i.e., no longer large compared with
unity), a field theory description must be used.

It is easy to calculate the experimental observables from the GP
wave function $\Psi(x,t_f)$.  By performing a Fourier transform it is
easy
to calculate the fraction $p_{nk}(t_f)$ of atoms associated with each
sharply peaked momentum component $n$.  The degree of
interband non-adiabaticity $f_{{\mathrm{nonad}}}$ is determined
experimentally by the amplitude of the oscillations in $p_{0k}(t_f)$
versus $t_f$:
\begin{equation}
   f_{{\mathrm{nonad}}} = p_{0k,{\mathrm{max}}} - p_{0k,{\mathrm{min}}}
   \,, \label{Eq.3}
\end{equation}
where $p_{0k,{\mathrm{max}}}$ and $p_{0k,{\mathrm{min}}}$ are the
maximum and minimum values of $p_{nk}(t_f)$ for $n=0$
over an interval $t_f-t_r
= t_{{\mathrm{hold}}}$ which is chosen to be long enough to
include at least one oscillation of $p_{0k}(t_f)$.  If there is no
interband excitation, then $f_{{\mathrm{nonad}}}=0$.

\begin{figure}
   \centering
   \includegraphics[scale=0.3]{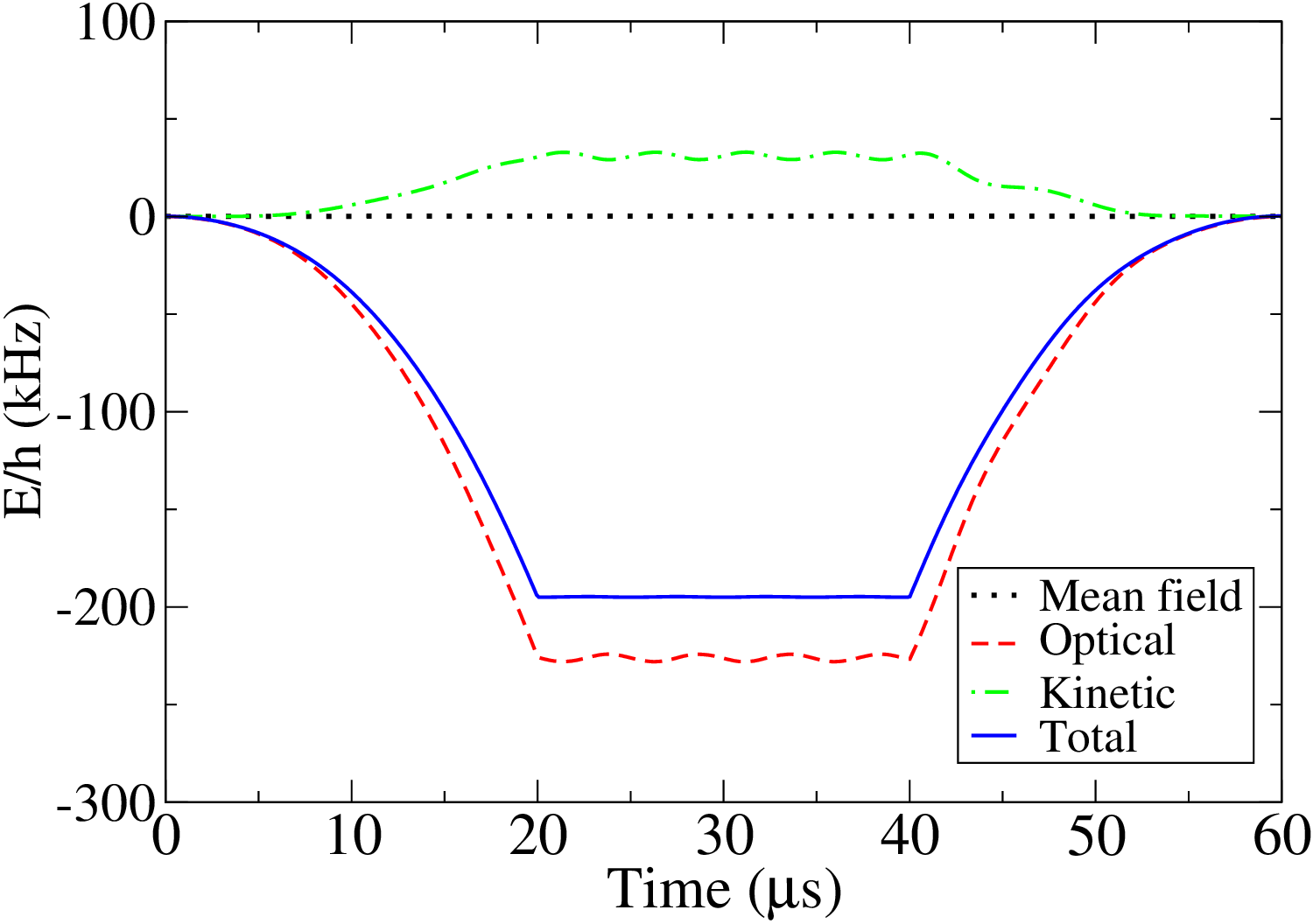}
   \caption{Kinetic, optical, mean-field and total energies versus
   time starting from a condensate with $10^4$ Na atoms in a trap 
   with frequencies of  84 Hz, 59.4 Hz, and 42 Hz in the respective 
   $x$, $y$, and $z$ directions.  A 20 $\mu$s linear ramp is applied
   ending at a latttice depth of $V_0=14
   E_R$ followed by a 20 $\mu$s hold time and a 20 $\mu$s linear
   ramp down.}
   \label{fig1}
\end{figure}

Figure~\ref{fig1} shows various components of the BEC energy as they
evolve in time when an optical potential with $2\pi/k=$ 590 nm, and
$V_0 = 14 E_R$, where the recoil energy $E_R$ is defined as
$E_R=\hbar^2 k^2/2m$, is turned on linearly over 20 $\mu$s, held
constant for 20 $\mu$s, and finally turned off over 20 $\mu$s.  The
trap with frequencies $\nu_x=84$ Hz, $\nu_y=59.4$ Hz, and $\nu_z=42$
Hz holds $10^4$ Na $F=1,M=-1$ atoms, for which $a=2.8$ nm and
$E_R/h=25$ kHz.  Notice that the total energy of the system is constant
when the potentials do not depend on time.  The mean field energy
remains negligible in comparison to the kinetic and potential energies
that result from turning on the optical potential.  Consequently,
calculations with 10,000 and 1$\times 10^6$ atoms are
indistinguishable on the scale of Fig.~\ref{fig1}.  We also checked that
on these short timescales the calculated nonadiabaticities from both
3D and 1D versions of the GP equations are indistinguishable.  Thus,
1D linear Schr{\" o}dinger dynamics should provide an excellent
approximation to the nonadiabatic dynamics.

\subsection{Unit-cell Model}

We only consider relatively short time scales $t_r$ of lattice loading
so that mean-field interactions do not play an important role, as shown 
in Section~\ref{sec.mf}.  We
assume $t_r$ is short compared with the nonlinear time
$t_{\mathrm{NL}} \equiv m/(4\pi\hbar a|\Psi_{\mathrm{peak}}|^2) =
\hbar /\mu$, where $|\Psi_{\mathrm{peak}}|^2$ is the peak value of
$|\Psi|^2$ at $t=0$, and $\mu$ is the chemical potential.  Consequently,
we can develop a simple {\em linear} `unit cell' model of the loading process
that is in excellent agreement with the full GP dynamics.  In the unit
cell model, the natural units of length, energy, and time are $1/k$,
the recoil energy $E_R=\hbar^2 k^2/2m$, and a characteristic lattice 
band excitation time $t_{\mathrm{band}} = 2\pi/\Delta \omega$ respectively.

\begin{figure}
   \centering
   \includegraphics[scale=0.33]{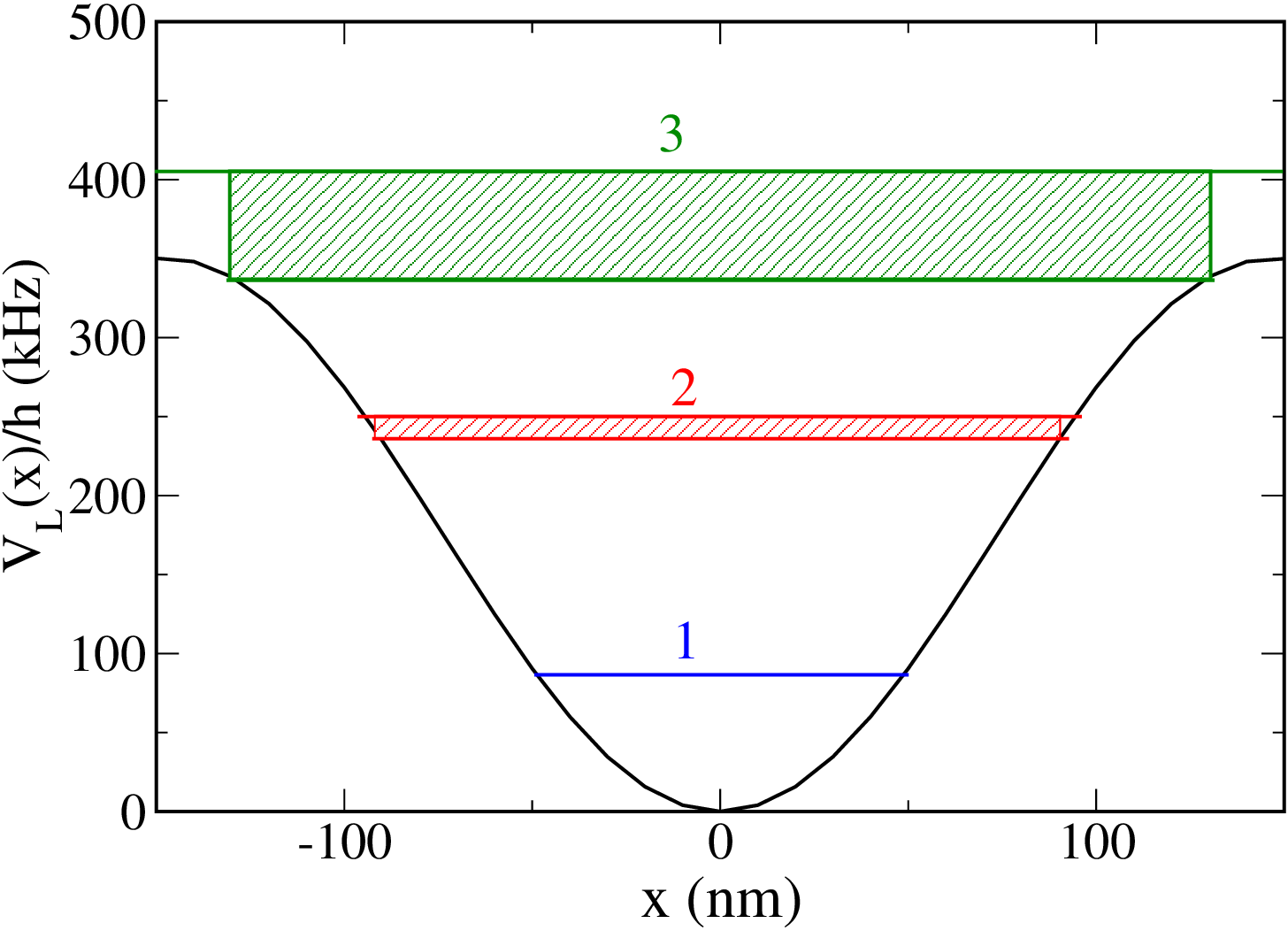}
   \caption{Unit cell lattice potential $V_L(x)$ versus $x$, showing
   the energy spread of the first three energy bands for a
   Na lattice with $2\pi/k=590$ nm and $V_0 = 14E_R$, where $E_R/h =
   25$ kHz.  \\}
   \label{fig2}
\end{figure}

\begin{figure}
   \centering
   \includegraphics[scale=0.27]{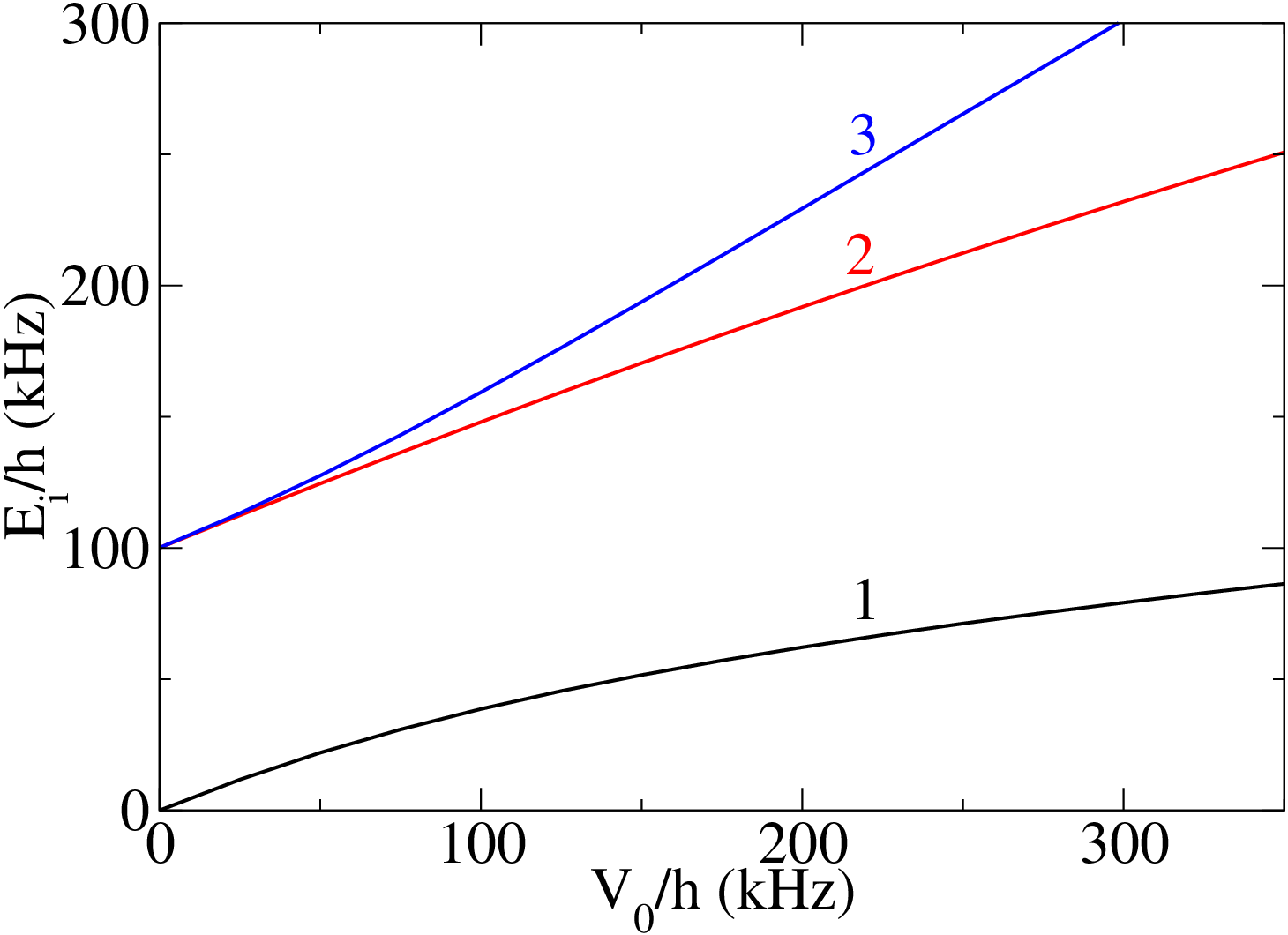}
   \caption{Energies of the first three $q=0$ levels versus lattice depth 
   $V_0/h$ for the unit cell shown in Fig.~\ref{fig1} .
   The second and third levels are degenerate with an energy $4E_R$ in
   free space, but split into symmetric ($E_3)$ and antisymmetric
   ($E_2$) states with increasing lattice depth.  Turning on the
   lattice rapidly couples the symmetric $i=1$ ground and $i=3$
   levels, but not the $i=2$ one.  \\}
   \label{fig3}
\end{figure}

Fig.~\ref{fig2} shows the Wigner-Seitz primitive unit cell of an
infinite optical lattice of depth $V_0 = 14 E_R$.  Solving the
Schrodinger equation for the potential $-V_0 \cos^2{kx}$ with periodic
boundary conditions on the interval $- \pi/2k \leq x \leq \pi/2k$
defines a series of eigenvalues $E_i(V_0)$ and eigenfunctions
$|i,V_0\rangle$ that describe the unit cell wave function for the
$q=0$ edge of each band $i$, where $q$ is the lattice momentum~\cite{MatEq}.  
The boxes in Fig.~\ref{fig2} indicate the width of the energy
bands.  The width of the lowest band is too narrow to be observed on
the energy scale of the figure.  The width of the third band extends
beyond the top of the lattice potential.  Fig.~\ref{fig3} shows the
energies $E_i$ for the $q=0$ levels for the 
first three lattice bands as the lattice depth $V_0$ is increased.  Since 
symmetry considerations only allow excitation of odd numbered bands from
the first $n=1$ band during lattice turn on, the relevant trap 
excitation parameter $\Delta \omega$ is $(E_3-E_1)/\hbar$.  For
the example in Fig.~\ref{fig2}, $\Delta \omega = 2\pi(250 \,\mathrm{kHz})$ 
and $t_{\mathrm{band}}=4$ $\mu$s.

An adiabatic basis set expansion can be used to calculate the
time-dependent amplitudes $a_j(t)$ of the instantaneous
eigenvectors during the dynamics.  The basis set expansion is
taken in the form,
\begin{equation}
   \Psi(t)= \sum_{j}a_j(t) e^{-i\int_0^t E_j(t')dt'/\hbar} | j(t)
   \rangle \,,
   \label{adiab_expansion}
\end{equation}
where $| j(t) \rangle$, $j=1,2, \ldots$, is the instantaneous
eigenvector with
energy eigenvalue $E_j(t)$ corresponding to an optical potential
strength $S(t) V_0$.   The nonadiabatic dynamics is then calculated
from the coupled set of equations,
\begin{equation}
   \dot{a_i} = \sum_{j \ne i} a_j(t)\frac{\langle i(t)|\dot{H}|
   j(t)\rangle}{E_i(t)-E_j(t)}\nonumber e^{i(\alpha_j(t)-\alpha_i(t))}
   \,.  \label{adiab_eqs}
\end{equation}
where we have defined the coefficient $\alpha_j(t)=-\int_0^t
E_j(t')dt'/\hbar$ to simplify the notation.  The even index and
odd index eigenstates are not coupled because the potential and its
time derivative are an even function of $x$ and the odd (even)
eigenstates are symmetric (antisymmetric).  We numerically
calculate $E_i(t)$, $E_j(t)$ and $\langle i(t) |\dot{H}(t) | j(t)
\rangle$, and solve the coupled differential equations by using a
variable step size integrator.  We will call this method the
converged unit cell model.

We have verified by numerical calculations that a truncated basis set
expansion with the lowest two terms is a good approximation unless
the ramp time $t_r$ becomes comparable to $t_{\mathrm{band}}$ or less:
\begin{equation}
  \Psi(t) = a_1(t) e^{i\alpha_1(t)} | 1(t)\rangle + a_3(t)
  e^{i\alpha_3(t)} | 3(t)\rangle \,,
  \label{Eq.4}
\end{equation}
This leads to the following simplified 2-equation unit cell model:
\begin{equation}
   \dot{a_1}=a_3(t)\frac{\langle 1(t) | \dot{H} |
   3(t)\rangle}{E_1(t)-E_3(t)} e^{i(\alpha_3(t)-\alpha_1(t))} \,,
   \label{Eq.5}
\end{equation}
\begin{equation}
   \dot{a_3}=a_1(t) \frac{\langle 3(t) | \dot{H} |
   1(t)\rangle}{E_3(t)-E_1(t)} e^{i(\alpha_1(t)-\alpha_3(t))} \,.
   \label{Eq.6}
\end{equation}

An even simpler one-equation approximation, which is acceptable for
sufficiently large $t_r$, is to set $a_1=1+i0$ and solve for
$a_3(t)$ by integrating Eq.~(\ref{Eq.6}), so that
\begin{equation}
   a_3(t) = \int_0^t \frac{\langle 3(t') | \dot{H} |
   1(t')\rangle}{E_3(t')-E_1(t')} e^{i(\alpha_1(t')-\alpha_3(t'))} dt'
   \,.
   \label{1_eq}
\end{equation}
We will call this the one-equation unit cell model.

The fraction of the total number of atoms in the wave packet with mean
momentum $nk$, $p_{nk}(t)$, can be calculated in terms of the
amplitudes $ a_j(t)$:
\begin{eqnarray}
   p_{nk}(t) =| \langle 2nk| \Psi(t) \rangle |^2 &=& \left | \sum_{j}
   a_j(t)\langle 2nk| j(t) \rangle e^{i\alpha_j(t)} \right |^2
   \nonumber \\
   \approx |a_1(t)\langle 2nk| 1(t) \rangle e^{i\alpha_1(t)} &+&
   a_3(t)\langle 2nk| 3(t) \rangle e^{i\alpha_3(t)}|^2 .
   \label{Eq.7}
\end{eqnarray}
When $n>0$, the total population with magnitude of momentum $|nk|$ is
\begin{equation}
  p_{|nk|}(t) = p_{nk}(t)+p_{-nk}(t) \,.
  \label{Eq.7b}
\end{equation}

The observable $f_{{\mathrm{nonad}}}$ in Eq.~(\ref{Eq.3}) can be
calculated from $p_{0k}(t)$.  Using the one-equation model
of Eq.~(\ref{1_eq}), where $a_1=1$, $f_{{\mathrm{nonad}}}$ simplifies to
\begin{equation}
   f_{{\mathrm{nonad}}} =
   4 \langle 0k | 1 \rangle \langle 0k | 3 \rangle |a_3| \,.
   \label{Eq.8}
\end{equation}
The expression in Eq.~(\ref{Eq.8}) is evaluated for $t>t_r$, where none of 
$V_L$, $|1 \rangle$, $|3 \rangle$, or $|a_3|$ is changing in time.

The fidelity $F$ of the ramp can be defined as the
fraction of population that is left in the ground state after the
ramp: $F=|a_1|^2$, where $a_1(t)$ is evaluated for a time after $t_r$.
In the two-state model, $F=1-|a_3|^2$.  Using the result in
Eq.~(\ref{Eq.8}), we can get a relation between $F$ and
$f_{\mathrm{nonad}}$:
\begin{equation}
 F = 1 - \frac{1}{16 | \langle 0k | 1 \rangle \langle 0k | 3
\rangle|^2}  f^2_{{\mathrm{nonad}}}
 \label{fidelity}
\end{equation}
For the case of the $V_0=14E_R$ Na lattice discussed in the figures,
$F=1-0.28 |f_{\mathrm{nonad}}|^2$.

\begin{figure}
\centering
\includegraphics[scale=0.33]{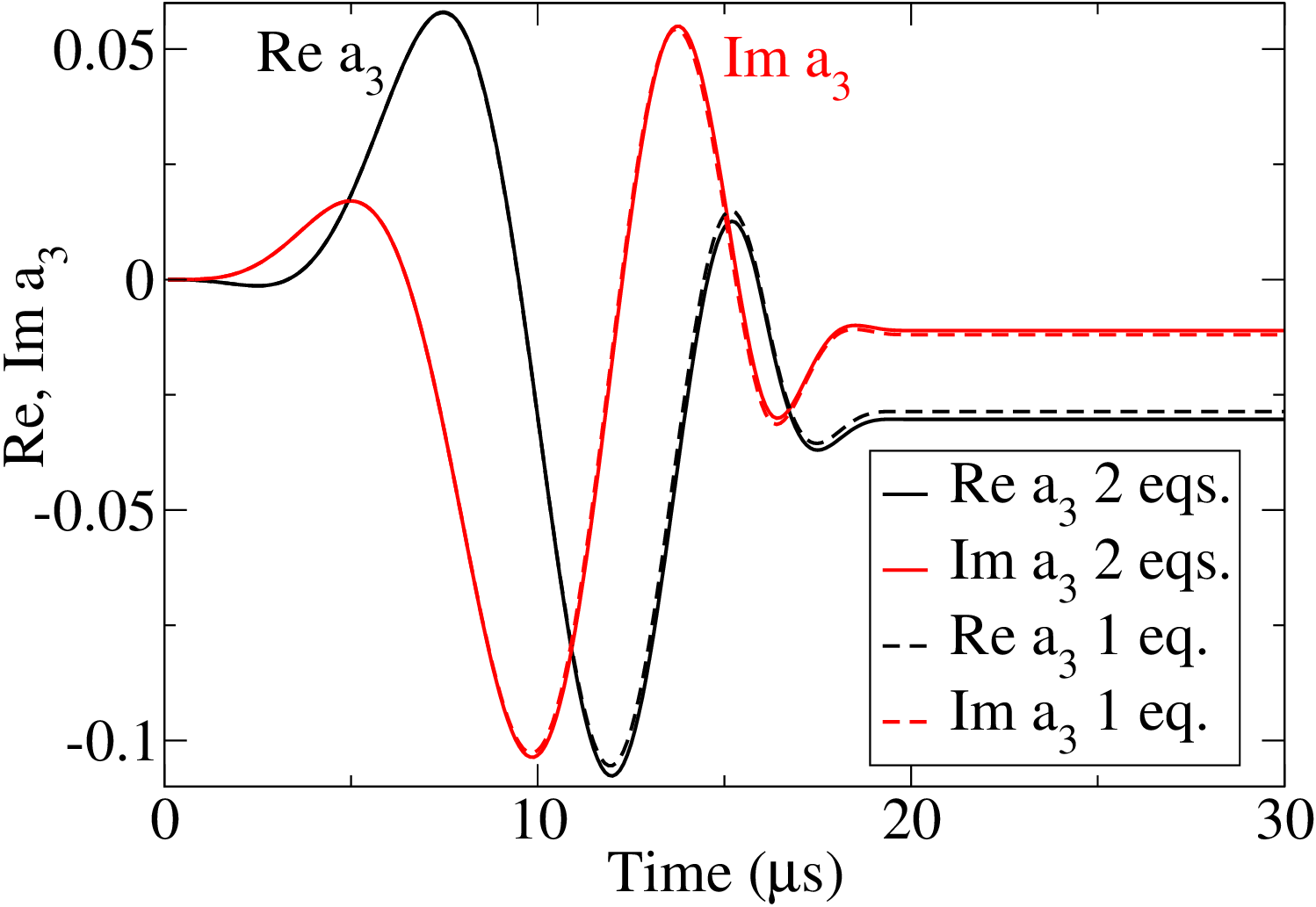}
\caption{Time evolution of the real and imaginary parts of
$a_3$ calculated with the $14 E_R$ deep, 20 $\mu$s cubic ramp shown
in Fig.~\ref{fig.ramp}.  The results are shown for the two equation
model and the one equation model. Eqs.~(\ref{Eq.5})-(\ref{Eq.6}) and
Eq.~(\ref{1_eq}) respectively. \\.} \label{fig4}
\end{figure}

Fig.~\ref{fig4} compares the numerical solutions for the two-equation
model of Eqs.~(\ref{Eq.5}) - (\ref{Eq.6}) and the one-equation model
of Eq.~(\ref{1_eq}) for a $V_0=14E_R$ lattice with the cubic ramp
function in Fig.~\ref{fig.ramp}.  Clearly, the simpler one-equation
model provides a very good approximation in this case.  Figure~\ref{fig4} 
shows that the degree of nonadiabaticity at the end of the ramp can be
much smaller than during the turn-on portion of the ramp.  This
suggests that we might be able to optimize the ramp so that the final
value of $|a_3|^2$ at the end of the ramp is small, even if it becomes
large during the ramp.

\begin{figure}
   \centering
   \includegraphics[scale=0.28]{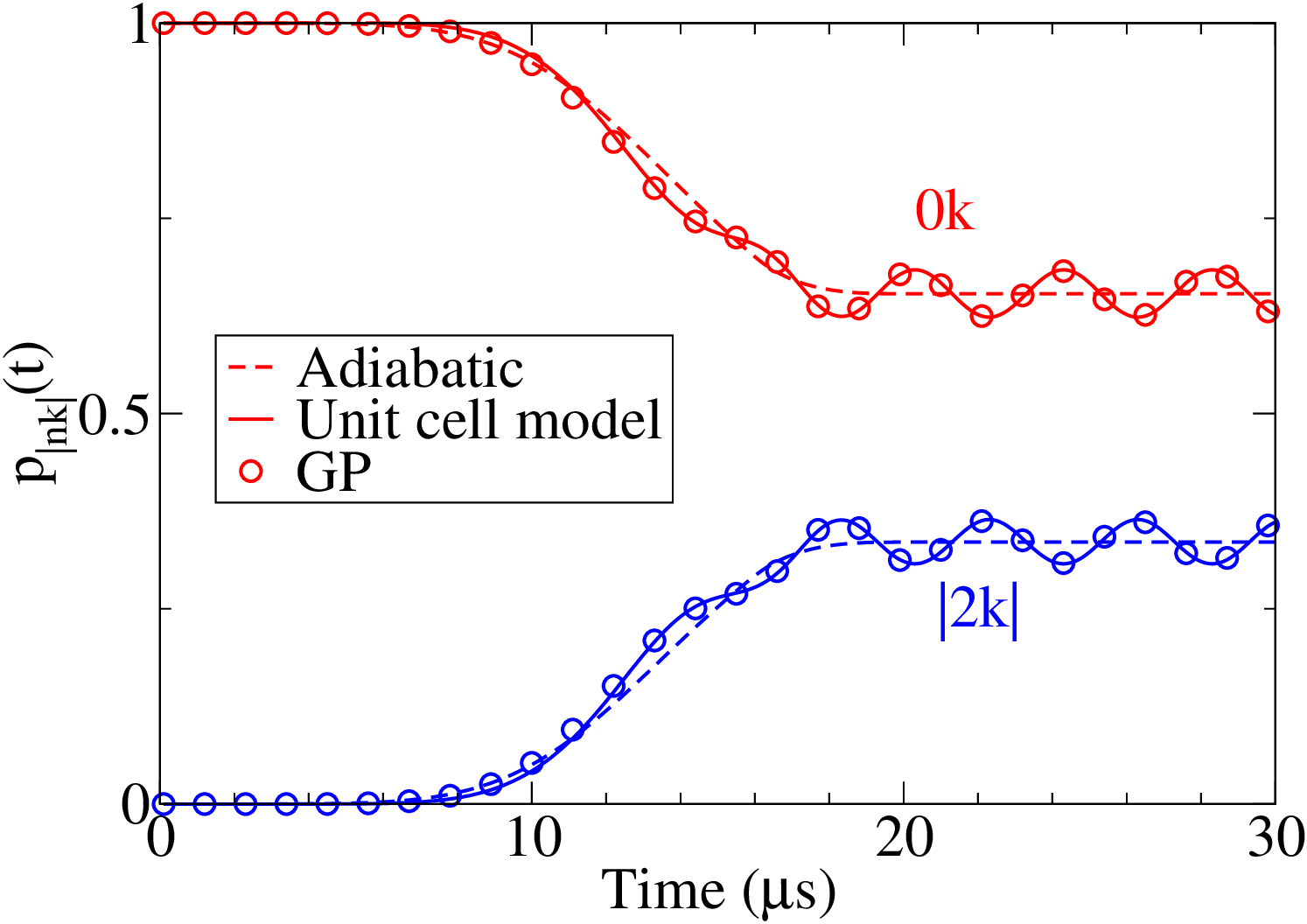}
   \caption{Population of the 0k and $|2k|$ momentum components versus
   time.  The optical lattice is turned on to a depth of $14 E_R$
   with the 20 $\mu$s cubic ramp shown in Fig.~\ref{fig.ramp}. \\}
   \label{fig5}
\end{figure}

Figure~\ref{fig5} shows the populations of the free space momentum
states, $p_{0k}(t)$ and $p_{|2k|}(t)$, for
several different calculations.  The $V_0=14E_R$ lattice is turned on
with the cubic ramp function in Fig.~\ref{fig.ramp}.  The figure shows the
momentum state populations that correspond to the adiabatic
eigenstates, $|\langle 2nk|j\rangle|^2$, as well as those obtained
with dynamical calculations based on the GP equation and the
two-equation unit cell model of Eqs.~(\ref{Eq.5}) - (\ref{Eq.6}).  The
figure demonstrates the very good agreement between the GP and unit
cell model.  The oscillations of the dynamical populations around the
adiabatic ones are due to nonadiabatic excitation due to the fast
ramp.  The amplitude of these oscillations correspond to
$f_{\mathrm{nonad}} = 0.07$ and a fidelity of $F=0.9985$.

\subsection{Optimization}

The oscillations evident in Fig.~\ref{fig5} are due to the small but
finite value of nonadiabatic excitation that reduces the fidelity $F$
of the ramp.  We will use an unconstrained nonlinear
optimization to minimize the degree of nonadiabatic excitation at the
end of the ramp and thus maximize the fidelity F. For this purpose we
use an $N$th order polynomial form for the ramp function $S(t)$, with
$S(0) = 0$ and $S(t_r) = 1$.  The form of $S(t)$ is given by
\begin{equation}
     S(t)=\sum_{i=0}^{N+1} c_i t^i \,,
     \label{eq:S(t)}
\end{equation}
where $c_{N+1} = (1 - \sum_{i=0}^{N} c_i t_r^i)/t_r^{N+1}$ insures
that $S(t_r) = 1$.  We then can use a standard computer code to
optimize the polynomial coefficients $c_i$, $i = 1,\ldots, N$ to give
the minimal $f_{{\mathrm{nonad}}}$ .  We can optimize
$f_{{\mathrm{nonad}}}$ calculated from the GP equation, or from the
one- or two-equation unit cell models.  We will use the GP equation,
since this allows us to work with very short ramps, where the
nonadiabaticity is large during the ramp.  We will use the simpler
models for interpreting the results.

\section{Results}

\subsection{Comparison to experiment}

In the experiment of Ref.~\cite{Denschlag02}, a BEC with $3 \times 10^6$ sodium
atoms in the 3S$_{1/2}$, $F = 1$, $m_F = -1$ state with no discernible
thermal component is prepared and held in a magnetic time-orbiting
potential (TOP) trap with trapping frequencies in the $x$, $y$ and $z$
directions of 27 Hz, 19 Hz and 13.5 Hz.  After production of the condensate,
an optical lattice consisting of two counter-propagating laser beams
along the x-direction is turned on.  The lattice beams are detuned
about 60 GHz to the blue of the sodium D2 line
at 589 nm.  The spontaneous emission rate is negligible on the
timescale of the experiments.  The polarization is linear and parallel
to the rotation axis of the TOP trap bias field (the y-axis).  The
condensate is located in the focus of the beams, which have a $1/e^2$ beam
diameter of about 600 $\mu$m.

\begin{figure}
   \centering
   \includegraphics[scale=0.33]{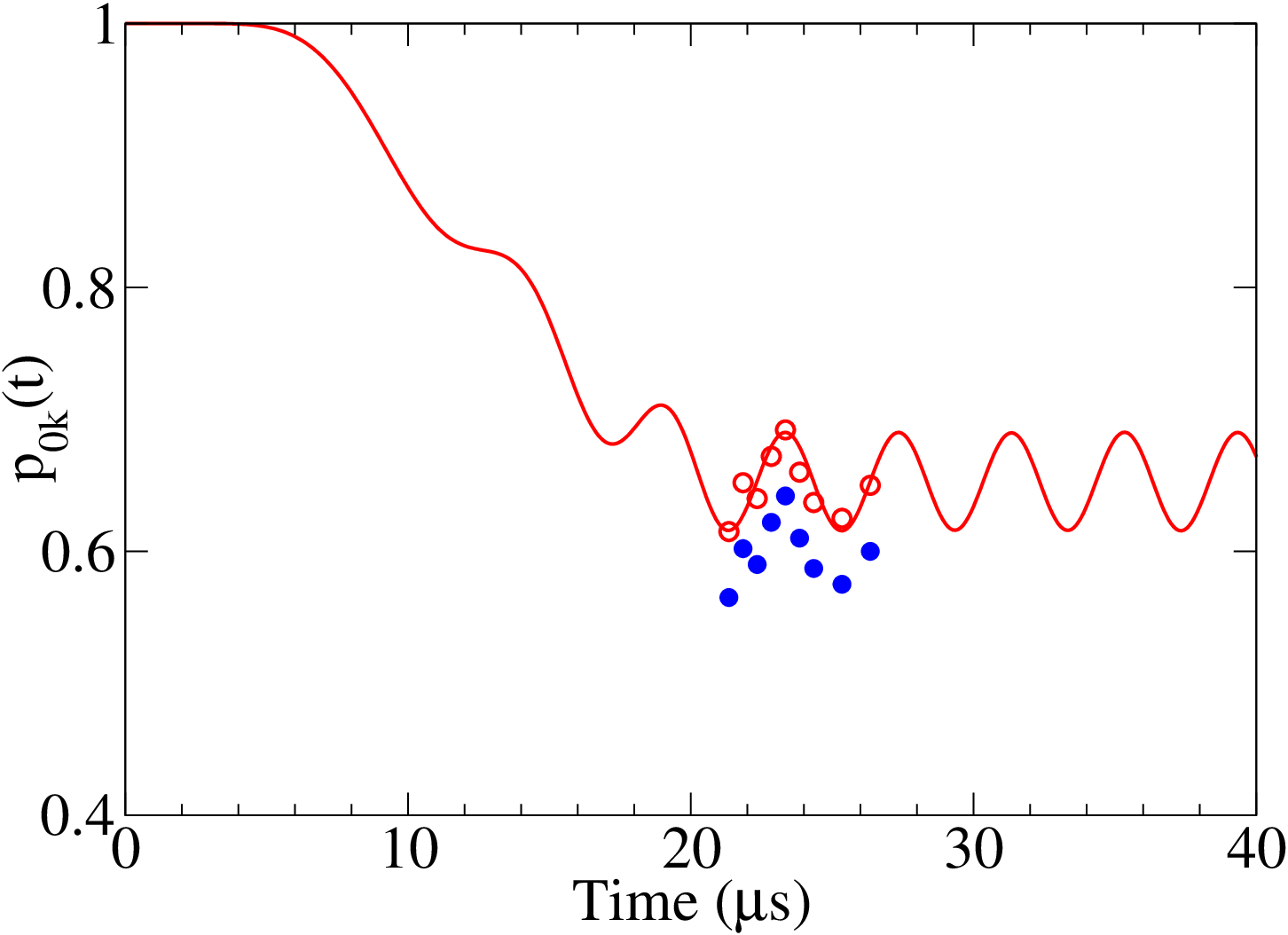}
   \caption{Comparison of theory and NIST experimental data (closed
   circles) \cite{Denschlag02} on the population of the 0$k$ momentum
   component versus time for a 20 $\mu$s ramp of a $V_0=14E_R$ deep
   optical lattice.  The open circles shows data shifted to have the
   same peak magnitude as that of the calculation.  \\
   \bigskip}
   \label{fig6}
\end{figure}

In order to directly measure the adiabaticity, i.e., the efficiency of
transfer into the lattice ground state, an experiment was carried out
in which the intensity of the optical potential was ramped up to a
stationary lattice potential strength $V_0 \approx 14E_R$ \cite{Denschlag02}.  The
BEC was held in the lattice for a time $t_{\mathrm{hold}}$, typically
between 0 and 10 $\mu$s, before suddenly switching off the light.  The
plane-wave decomposition of the lattice wave function at the switch-off time $t$
depends on the previous history of the ramp prior to time $t$.  The 
populations of the various momentum components at $t$ is measured by allowing
the atomic cloud to expand after the switch-off, since each $2nk$ momentum
component eventually separates from the others in an individual cloud
that can be imaged.  Figure \ref{fig6} shows calculated and observed
oscillations for the population of the $0k$ component following a ramp
of 20 $\mu$s.  This beating signal has a small amplitude,
indicating that most of the population was in the lowest band.
However, if only the ground state were populated, there would have
been no beating at all.

Figure \ref{fig6} also shows our calculation of $p_{0k}(t)$ using the
GP equation with the experimental lattice potential having the ramp
shown in Fig.~\ref{fig.ramp}.  The figure shows the original data of
Ref.~\cite{Denschlag02} as well as data that has been shifted in
phase and magnitude to coincide with the calculation.  These
adjustments are consistent with an experimental uncertainty in the
lattice depth $V_0$ of about 10\% and the uncertainty in the ramp
function.  Even without the adjustments, it is clear that our
calculation gives the right magnitude of the effect of nonadiabatic
dynamics as measured by the oscillations in the signal after $t_r$.

\subsection{Ramp Optimization Results}

Although the degree of nonadiabaticity in Fig.~\ref{fig6} is small, it
can be made even smaller by optimizing the ramp turn on function.  
We used a numerical optimization algorithm to
minimize $f_{{\mathrm{nonad}}}$ as defined in Eq.~(\ref{Eq.3}) versus
the polynomial fit parameters in Eq.~(\ref{eq:S(t)}).  We calculated the
dynamics with the GP equation, the one-equation unit cell model, the
two-equation unit cell model, and the converged unit cell model.  The
one-equation unit cell model works very well except for short ramps
with $t_r$ on the order of  $t_{\mathrm{band}}$.  Only the GP and the
converged unit cell model work well in this limit, due to excitation
of higher bands that are not modeled within the one- or two-equation
unit cell model.

\begin{figure}
   \centering
   \includegraphics[scale=0.33]{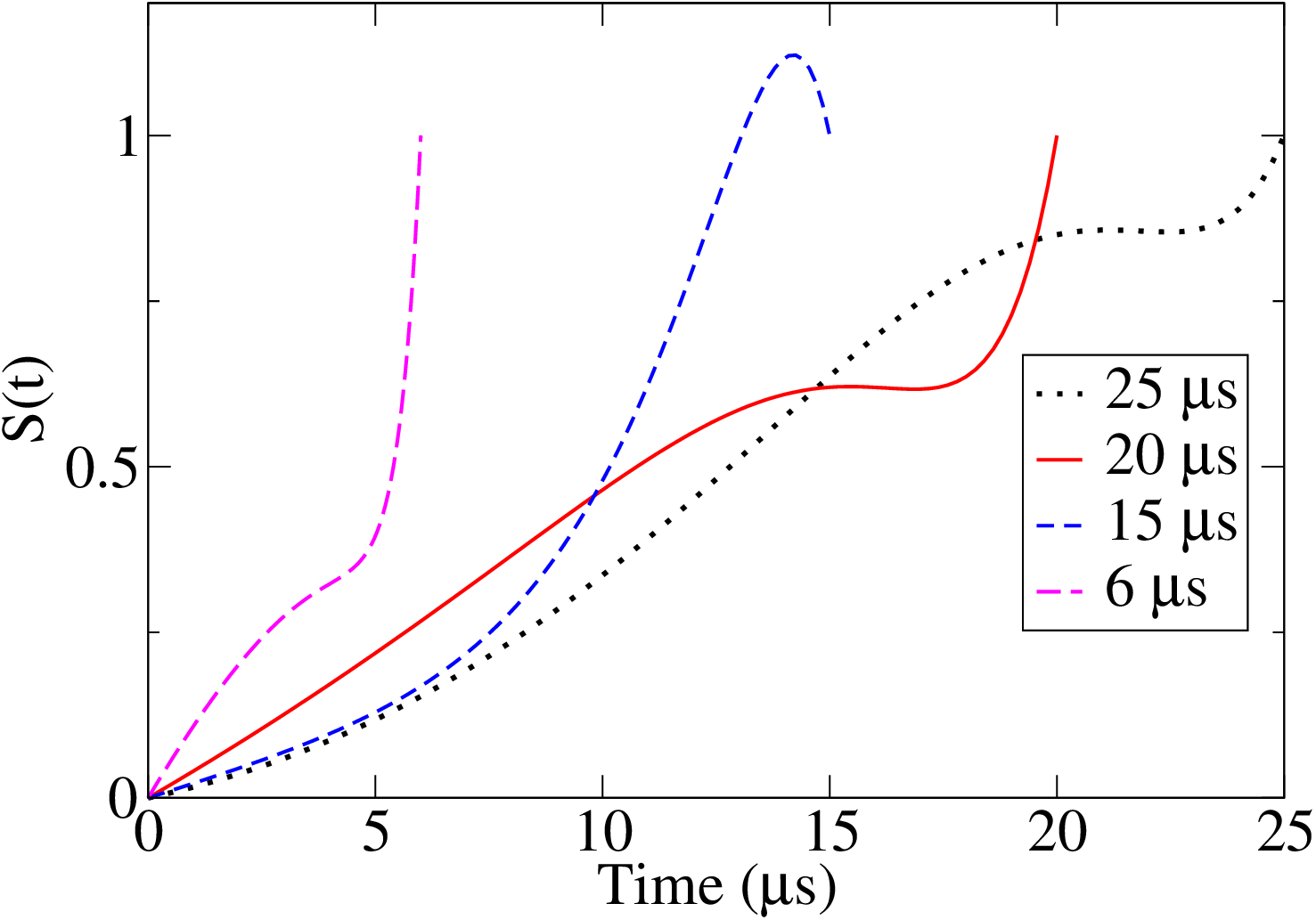}
   \caption{Optimal ramp functions for 6 $\mu$s, 15 $\mu$s, 20 $\mu$s, 
   and 25 $\mu$s ramps for  a $14E_R$ lattice. \\
   \bigskip}
   \label{fig7}
\end{figure}

Figure \ref{fig7} shows the optimal ramps obtained for several different
ramp durations, using $N=12$ in Eq.~(\ref{eq:S(t)}).  Clearly, the optimal 
ramp shape  depends strongly on the
ramp duration.  For very short $t_r$, the ramps are not robust in the
sense that slight deviations from the optimal ramp shape significantly
lowers the fidelity.  Moreover, different optimal ramps can be obtained
upon starting with varying initial functions for $S(t)$.  Thus, the
optimization functional being minimized can have several local minima
so that several different optimized functions can have similar high 
fidelities.

\begin{figure}
   \centering
   \includegraphics[scale=0.33]{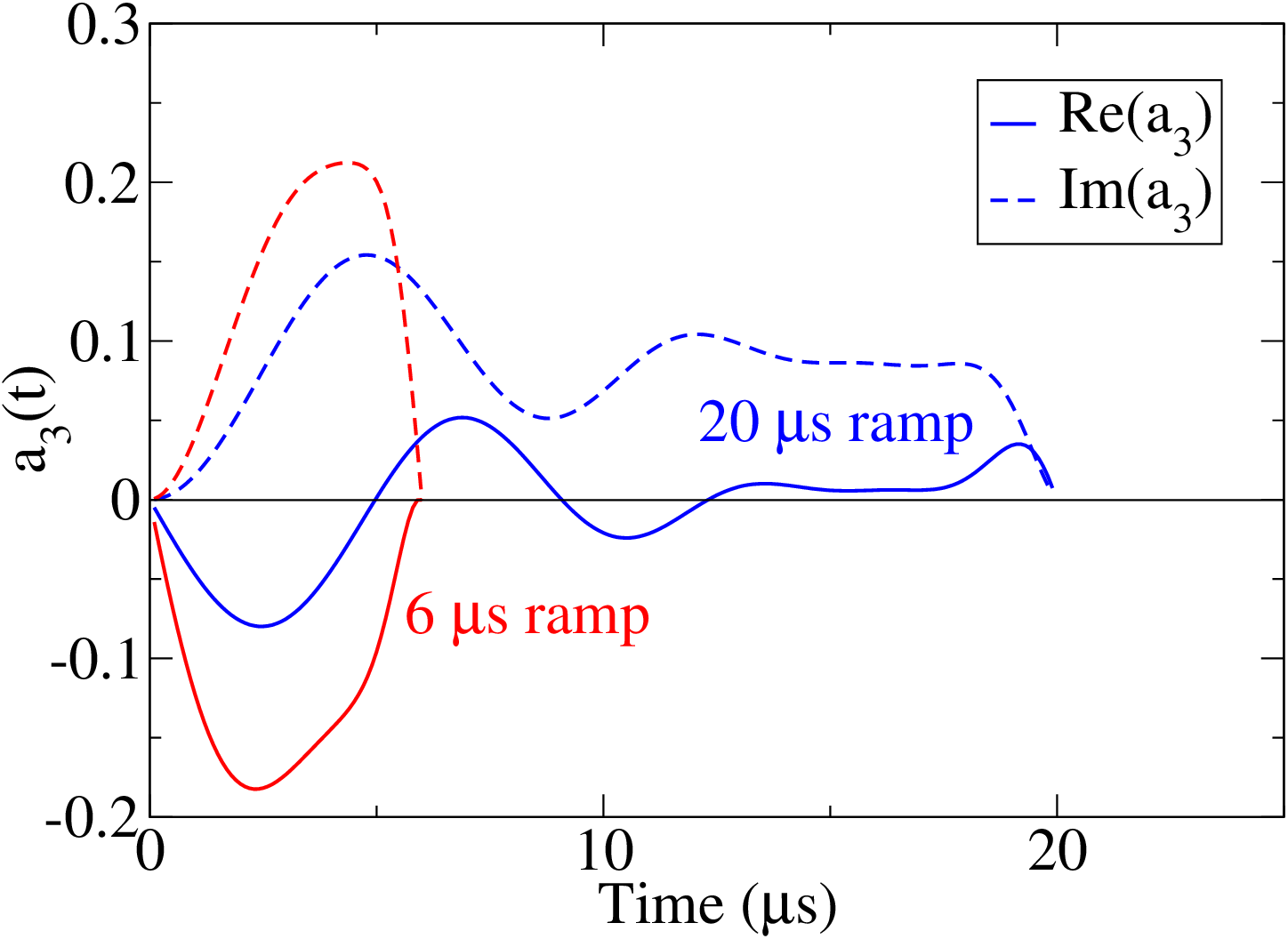}
   \caption{Real and imaginary parts of the amplitude $a_3(t)$ versus
   time for 6 and 20 $\mu$s optimal ramps, showing strong nonadiabatic
   behavior as a function of time, but ending at $t_r$ with very small
   excitation. \\}
   \label{fig8}
\end{figure}

Figure \ref{fig8} shows $a_3(t)$ versus time for the optimal 6 $\mu$s and 20
$\mu$s ramps in Fig.~\ref{fig7}.  In both cases, there is clearly
significant excitation during the ramp, however, at the end of the
ramps, the excitation is nearly zero.  Hence, the dynamics is clearly
nonadiabatic for both ramps, but at the final time $t_r$, there is
very little excitation, i.e., high fidelity is achieved for both
ramps.  With the optimized ramps, the fidelity obtained is much higher
than that obtained in the NIST experiment, even with the 6 $\mu$s
ramp.  However, for the optimal 6 $\mu$s ramp the $a_3$ coefficient
goes to zero at the end of the ramp with a high slope, indicating that
the fidelity will not be robust with respect to small changes in the
ramp parameters.  Also, for small $t_r$, the optimized ramp functions
can vary dramatically and suddenly with time and therefore may be
difficult to achieve experimentally.

\begin{figure}
   \centering
   \includegraphics[scale=0.30]{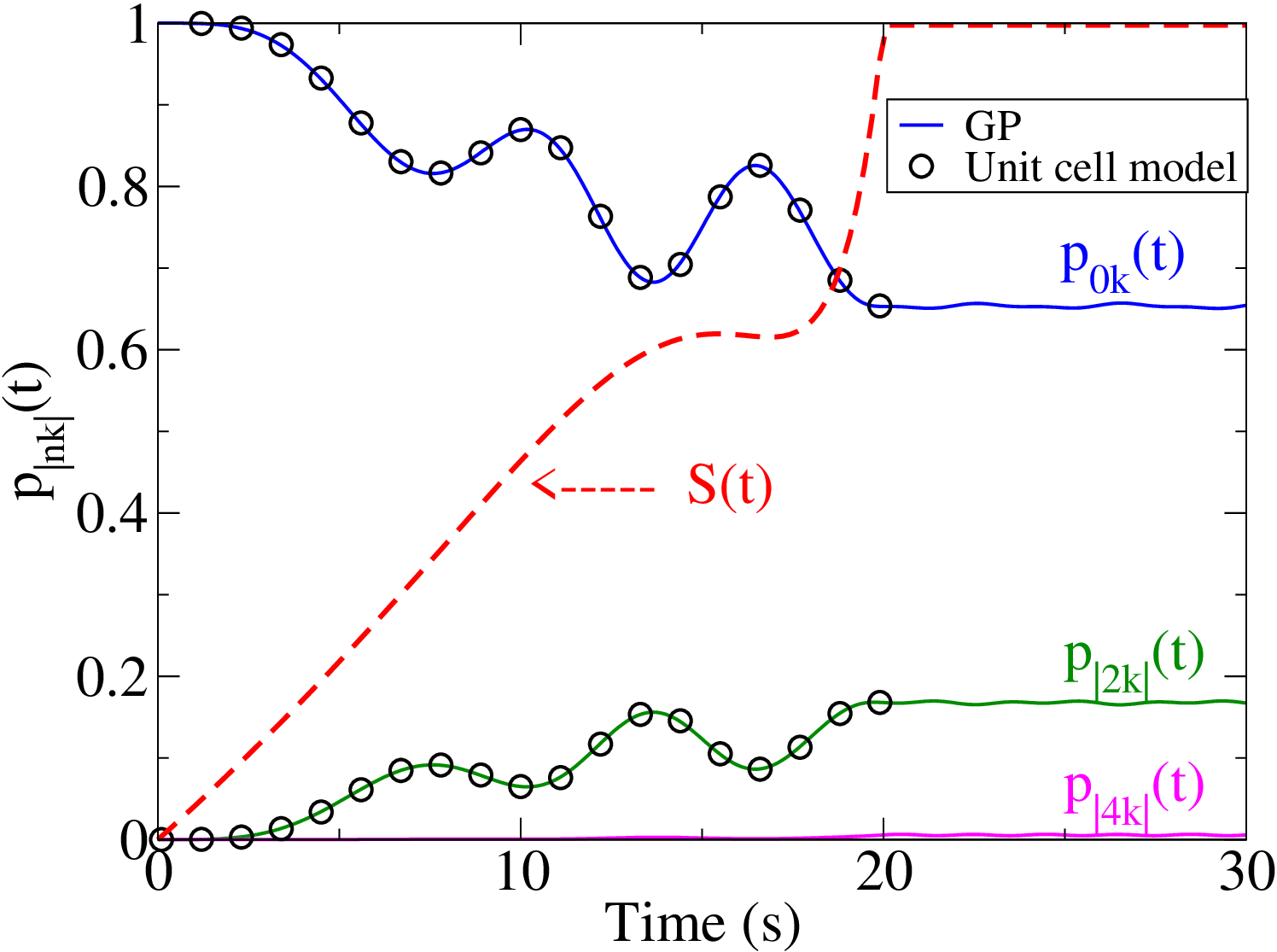}
   \caption{GP and unit cell populations versus time for
   optimal 20 $\mu$s ramp.  The fidelity is 0.999999. \\
   \bigskip}
   \label{fig9}
\end{figure}

\begin{figure}
   \centering
   \includegraphics[scale=0.28]{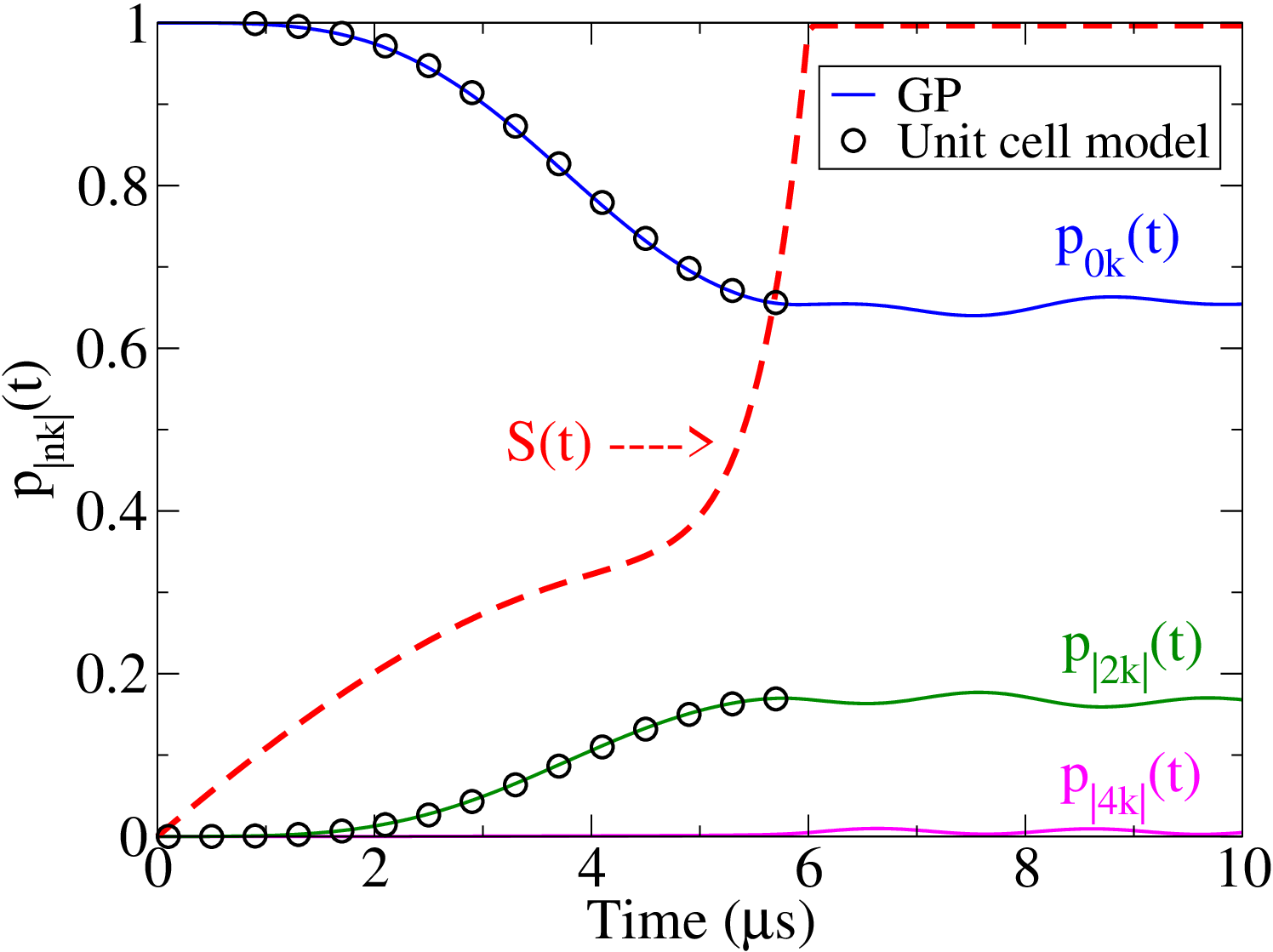}
   \caption{GP and unit cell populations versus time for
   optimal 6 $\mu$s ramp.  The fidelity is 0.9996. \\}
   \label{fig10}
\end{figure}

Figures \ref{fig9} and \ref{fig10} show the calculated $0k$ and $|2k|$
populations versus time as determined using the full GP equation and
the converged unit cell model for the optimal 20 and 6 $\mu$s ramps
respectively.  The $S(t)$ ramp functions are also shown in the
figures.  The converged unit cell model reproduces the GP results very
accurately, even for the 6 $\mu$s ramp.  For the 6 $\mu$s ramp, there
is already a small 4$k$ population component present.  Nevertheless,
the fidelity remains high.

\section{Summary}

We calculated the dynamics of a BEC upon turning on a one-dimensional
optical potential.  Good agreement with experimental data has been
obtained.  We showed that the optimal ramp function $S(t)$, with
$S(0)=0$ and $S(t_r)=1$, for turning on a one-dimensional optical
lattice potential, $V_L(x,t) = S(t) V_0
\cos^2(kx)$, can be chosen to minimize the interband nonadiabaticity.
A simple unit cell model that is in excellent agreement with the full
calculations has been developed for cases when the turn-on time $t_r$
is relatively long.  An optimized lattice turn-on with very low
nonadiabaticity can be chosen even for short loading times $t_r$
comparable to $2\pi/\Delta \omega$, where $\hbar \Delta \omega$ is the
band excitation energy.  

Even if the loading of a BEC into the lattice without causing
inter-band excitation is readily achievable, as shown here, unless
one switches on an optical lattice very slowly, the optical
lattice causes a spatially varying phase to accumulate across the
condensate due to intra-band excitation.  Ref.~\cite{Band_02} has
shown analytically and numerically that a cancellation of this
effect is possible by appropriately adjusting the harmonic trap
force-constant of the external magnetic trap that confines that
BEC, thereby facilitating quick loading of an optical lattice for
quantum computing purposes.

\begin{acknowledgments}
This work was supported in part by grants from the Office of Naval
Research, the U.S.-Israel Binational Science Foundation (grant
No.~2002147), the Israel Science Foundation for a Center of Excellence
(grant No.~8006/03), and the German Federal Ministry of Education and
Research (BMBF) through the DIP project.
\end{acknowledgments}


\begin{thebibliography}{}

\bibitem{Cirac&Zoller} J. I. Cirac and P. Zoller, Phys. Rev. Lett.
{\bf 74}, 4091 (1995).

\bibitem{Brune} M. Brune, P. Nussenzveig, F. Schmidt-Kaler, F. Bernardot, 
A. Maali, J. M. Raimond, and S. Haroche , Phys. Rev. Lett. {\bf 72},
3339 (1994).

\bibitem{Turchette} Q. A. Turchette, C. J. Hood, W. Lange, H. Mabuchi, 
and H. J. Kimble , Phys. Rev. Lett. {\bf 75}, 4710 (1995).

\bibitem{Gershenfeld} N. A. Gershenfeld and I. L. Chuang, Science
{\bf 275}, 350 (1997).

\bibitem{DiVi} D. P. DiVincenzo, J. of Magnetism and Magnetic Matls.
{\bf 200}, 202 (1999).

\bibitem{supercond} Y. Nakamura, Yu.  A. Pashkin, and J. S. Tsai,
Nature {\bf 398}, 786 (1999); J. E. Mooij, {\it et al.}, Science {\bf
285}, 1036 (1999); L. B. Ioffe, {\it et al.}, Nature {\bf 398}, 679 (1999).

\bibitem{Brennen_99} G. K. Brennen, C. M. Caves, P. S. 
Jessen, and I. H. Deutsch, Phys. Rev. Lett. {\bf 82}, 1060 (1999).

\bibitem{Jaksch_99} D. Jaksch, H.-J. Briegel, J. I. Cirac, C. W. Gardiner, 
and P. Zoller, Phys. Rev. Lett. {\bf 82}, 1975 (1999).

\bibitem{Jaksch_98} D. Jaksch, C. Bruder, J. I. Cirac, C. W. Gardiner, 
and P. Zoller, Phys. Rev. Lett. {\bf 81}, 3108 (1998).

\bibitem{Band_02}
S. E. Sklarz, I. Friedler, and D. J. Tannor, Y.B. Band, and C. J. Williams,
Phys. Rev. A{\bf 66}, 053620 (2002).

\bibitem{Orzel}
C.~Orzel, A. K. Tuchman, M. L. Fenselau, M. Yasuda, and M. A. Kasevich, 
Science {\bf 291}, 2386 (2001).

\bibitem{Greiner}
M. Greiner, O. Mandel, T. Esslinger, T. W. H{\" a}nsch, I. Bloch, 
Nature {\bf 415}, 39 (2002).

\bibitem{Denschlag02} J. Hecker Denschlag, J E Simsarian, 
H H{\" a}ffner, C McKenzie, A Browaeys, D Cho, K Helmerson, 
S L Rolston and W D Phillips, J. Phys. B{\bf 35}, 3095 (2002).

\bibitem{MatEq} Although we used a numerical solution for the potential, 
the analytic solution to the Mathieu equation could also be used.

\bibitem{Deng} L.Deng, E. W. Hagley, J. Wen, M. Trippenbach, Y. Band, 
P. S. Julienne, J. E. Simsarian, K. Helmerson, S. L. Rolston, W. D. Phillips, 
Nature {\bf 398}, 218 (1999).

\end{thebibliography}
\end{document}